\documentclass[twocolumn,prl,showpacs]{revtex4}
\usepackage{graphicx}

\begin{document}

\title{ Performance Limitations of Flat Histogram Methods and \\
        Optimality of Wang-Landau Sampling}

\author{P.~Dayal$^{1}$, S.~Trebst$^{1,2}$, S.~Wessel$^{1}$, D.~W\"urtz$^{1}$,
              M.~Troyer$^{1,2}$, S.~Sabhapandit$^{3}$, S.N.~Coppersmith$^{3}$}

\affiliation{$^{(1)}$Theoretische Physik, Eidgen\"ossische Technische Hochschule 
 Z\"urich, CH-8093 Z\"urich, Switzerland}
 \affiliation{$^{(2)}$ Computational Laboratory,  Eidgen\"ossische Technische Hochschule 
 Z\"urich, CH-8092 Z\"urich, Switzerland}
\affiliation{$^{(3)}$Department of Physics, University of Wisconsin, Madison, WI  53706 USA}
\date{\today}

\begin{abstract}
We determine the optimal scaling of local-update flat-histogram methods  with system size by using a perfect flat-histogram scheme based on the exact density of states of 2D Ising models.
The typical tunneling time needed to sample the entire bandwidth does not scale with the number of spins $N$ as the minimal $N^2$ of an unbiased random walk in energy space. While the scaling is power law for the ferromagnetic and fully frustrated Ising model, for the $\pm J$ nearest-neighbor spin glass the distribution of tunneling times is governed by a fat-tailed Fr\'echet extremal value distribution that obeys exponential scaling. We find that the Wang-Landau algorithm shows the same scaling as the perfect scheme and is thus optimal.
\end{abstract}

\pacs{02.70.Rr,75.10.Hk,64.60.Cn}

\maketitle

Monte Carlo methods are well-suited for the simulation of large many body problems,
since the complexity for a single Monte Carlo update step scales only polynomially and
often linearly in the system size, while the configuration space grows exponentially
with the system size. 
The performance of a Monte Carlo method is then determined by how many update
steps are needed to efficiently sample the configuration space. For second
order phase transitions in unfrustrated systems the problem of ``critical slowing down'' -- a rapid
divergence of the number of Monte Carlo steps needed to obtain a subsequent
uncorrelated configuration -- was solved more than a decade ago by cluster
update algorithms \cite{SwendsenWangWolf}. At first order phase transitions
and in systems with many local minima of the free energy such as frustrated
magnets or spin glasses, there is the similar problem of long tunneling times
between local minima. With energy barriers $\Delta E$ scaling linearly with
the linear system size $L$, the tunneling times $\tau$ at an inverse temperature
$\beta=1/k_B T$ scale exponentially with the system size,
$\tau \sim \exp(\beta\Delta E)\propto\exp({\rm const}\times L)$. 
Several methods were developed to overcome this tunneling problem, such as
the multicanonical method \cite{Multicanonical}, broad histograms \cite{Oliveira:96}, simulated and parallel tempering \cite{Tempering}, and Wang-Landau sampling \cite{WangLandau}.
The common aim of all these methods is to broaden the range of energies sampled within
Monte Carlo simulations from the sharply peaked distribution of canonical sampling
at fixed temperature in order to ease the tunneling through barriers. 

Ideally, all relevant energy levels are sampled equally often during a simulation, thus producing a ``flat histogram" in energy space. Some methods approach this goal by variations and generalizations of canonical distributions \cite{Multicanonical,Tempering}, while others \cite{Oliveira:96,WangLandau} discard the notion of temperature completely and instead are formulated in terms of the density of states.
With a probability $p(E)$ for a single configuration with energy $E$,
the probability of sampling an arbitrary configuration with energy $E$ is given as
$P_E=\rho(E) p(E)$, where the density of states $\rho(E)$ counts the number of states
with energy $E$.
Upon choosing $p(E)\propto1/\rho(E)$ instead of $p(E)\propto\exp(-\beta E)$
one obtains a constant probability $P_E$ for visiting each energy level $E$,
and hence a flat histogram.
Wang and Landau \cite{WangLandau}
proposed a simple and elegant flat histogram algorithm that
iteratively improves approximations to the initially unknown
density of states $\rho(E)$.
Once $\rho(E)$ is determined with sufficient accuracy, the Monte Carlo algorithm
just performs a random walk in energy space. 
Within two years of publication this algorithm has been applied
to a large number of problems \cite{Applications,Schulz:02,Yamaguchi:02} and extended
to quantum systems \cite{Troyer:03}.

In this Letter we investigate the performance of flat histogram algorithms in general, and the Wang-Landau algorithm in particular, for three systems  for which the density of states $\rho(E)$ is known exactly on finite two-dimensional (2D) lattices: the Ising ferromagnet as the simplest example,
the fully frustrated Ising model as a prototype for frustrated systems, and the $\pm J$ Ising spin glass.
For each of these models we construct a {\it perfect flat histogram method}
by simulating a random walk in configuration space where we employ the
{\it known density of states} for these models to set $p(E)\propto1/\rho(E)$. 

As a measure of performance we use the average tunneling time $\tau$ to get
from a ground state (lowest energy configuration) to an anti-ground state
(configuration of highest energy), which is the relevant time scale for
sampling the whole phase space \cite{footnote}.
Since the number of energy levels in a
$d$-dimensional system with linear size $L$ scales with the number of spins $N=L^d$,
the tunneling time for a pure random walk in energy space is
\begin{equation}
\tau\propto N^2=L^{2d}\;.
\label{Eq:PureRandomWalk}
\end{equation}
This scaling was in fact found for first order phase transitions \cite{Multicanonical,WangLandau}. 
However, none of the systems we study exhibit this scaling. While the ferromagnetic and fully-frustrated models exhibit power law scaling, for the spin glass the distribution of characteristic tunneling times is extremely broad and appears to diverge exponentially with system size.
For all three models the scaling of the performance of the Wang-Landau algorithm is the same as that of the perfect flat histogram method.

\begin{figure}
  \includegraphics[scale=0.35]{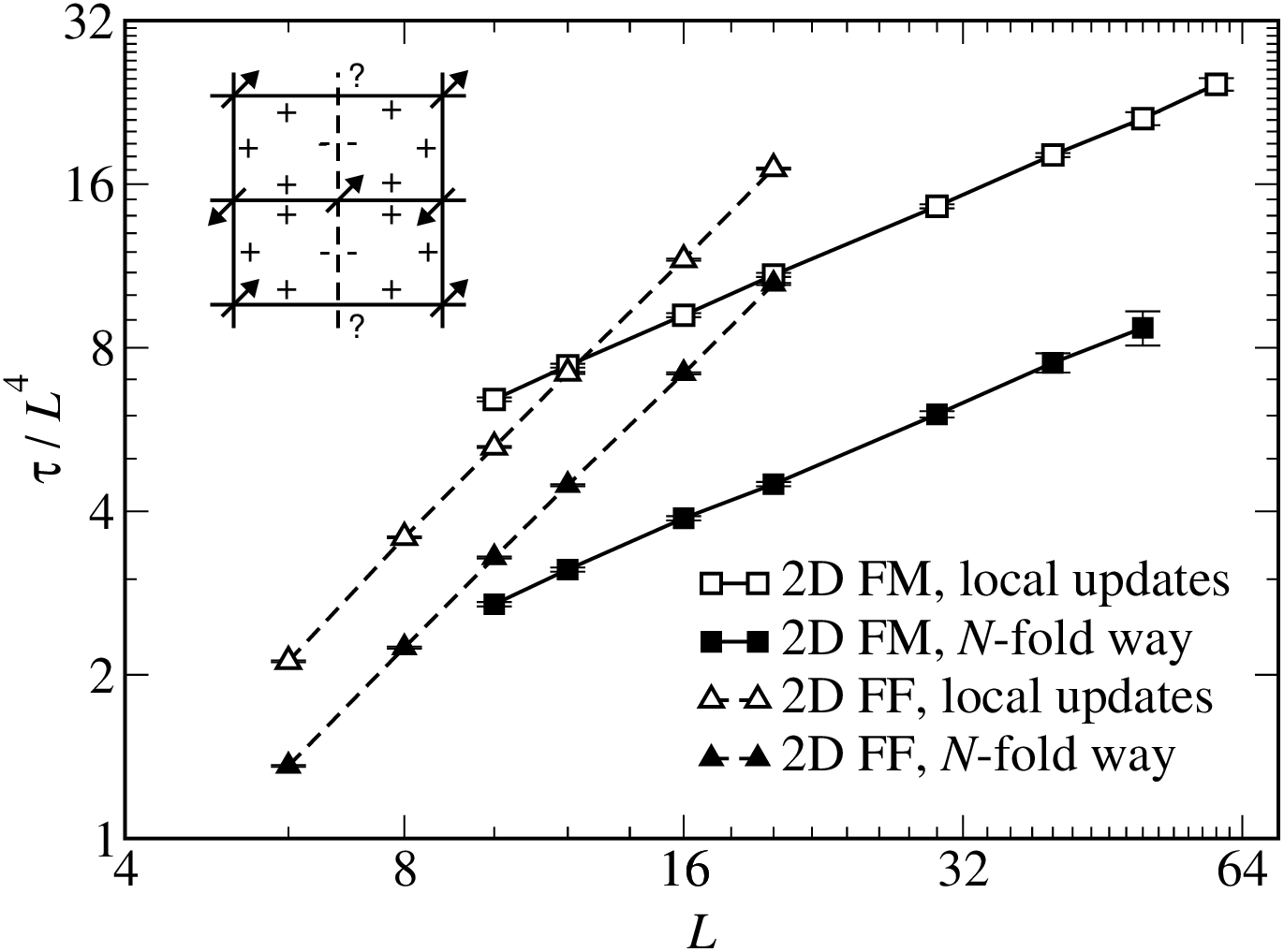}
  \caption{Scaling of tunneling times $\tau$ from the ground state to the
    anti-ground state as a function of system size $L$ for a perfect flat
    histogram method that samples using the exact density of states.
    Shown are results for the ferromagnet (FM) and fully frustrated (FF) 2D Ising 
    models with both local and N-fold way updates; the inset illustrates the 
    frustrated couplings.
    In all cases polynomial scaling $\tau \propto L^{2d+z}$ is found, with 
    $z_{\rm local}^{\rm FM}=0.743 \pm 0.007$, 
    $z_{N\rm -fold}^{\rm FM}=0.729\pm 0.011$,
    $z_{\rm local}^{\rm FF}=1.727\pm 0.004$, and 
    $z_{N\rm -fold}^{\rm FF}=1.692\pm 0.004$.}
  \label{Fig:Scaling_PureModels}
\end{figure}

We first look at homogeneous systems, with both ferromagnetic and fully frustrated couplings (see inset of Fig.~\ref{Fig:Scaling_PureModels}).
Exact densities of states $\rho(E)$ were calculated using the program of Beale \cite{Beale:96} for the ferromagnet and the algorithm of Saul and Kardar  \cite{Saul:94} for the fully frustrated Ising
model.
In  Fig.~\ref{Fig:Scaling_PureModels} the measured tunneling times are
plotted versus system size $L$ for the perfect flat histogram method
using both local and $N$-fold way updates.
Instead of the expected scaling Eq. (\ref{Eq:PureRandomWalk}),
we find a more rapid increase following $\tau \propto L^{2d+z}$ with
$ z_{\rm local}^{\rm FM}=0.743\pm0.007$ for the ferromagnetic (FM) model
and $z_{\rm local}^{\rm FF}=1.727\pm0.004$
for the fully frustrated (FF) model.
Frustration significantly increases the scaling exponent
but still conserves power law scaling.

\begin{figure}
  \includegraphics[scale=0.325]{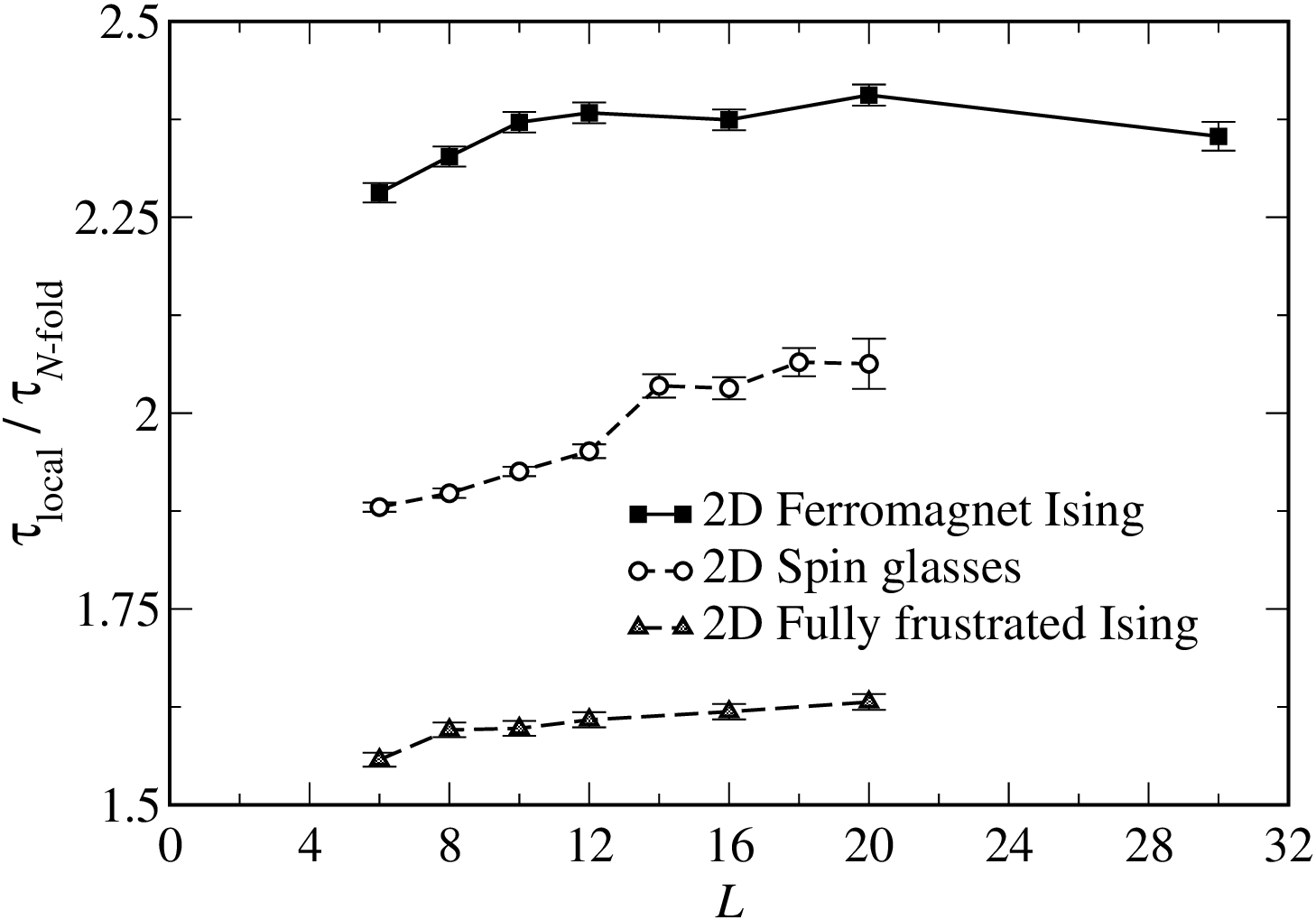}
  \caption{Scaling of the ratios of tunneling times measured using local and
    $N$-fold way updates. All simulation are done with single-spin flip updates
    weighted by the exact density of states. For each system size of the ferromagnet
    and the fully frustrated 2D Ising models 50,000 tunneling events and at least 25 tunneling 
    events for the 2D spin glass 1000 randomly generated realizations for
    $L = 6,8 \ldots 18$ and 350 realizations for $L$= 20 are used for averaging.}
  \label{Fig:Ratio_Local_NFoldWay}
\end{figure}

A previous study \cite{Schulz:02} suggests that $N$-fold way updates \cite{NFoldWay} speed up
Wang-Landau sampling.
We find identical scaling exponents for
$N$-fold way and local updates within our error bars, see Fig.~\ref{Fig:Scaling_PureModels},
implying that any performance improvement remains constant with system size.
As can be seen from Fig.~\ref{Fig:Ratio_Local_NFoldWay}, $N$-fold way updates
reduce the tunneling time by roughly a factor of 2, independent of system size. 
In practice, the reduction of tunneling times is offset by the added expense
of $N$-fold way updates.
The slight decrease of tunneling ratios in the fully frustrated model
compared to the ferromagnetic model can be explained by the increasing degeneracy
of ground states in the fully frustrated model,
which decreases the ground state lifetime and
makes $N$-fold way updates less effective. 


\begin{figure}
  \includegraphics[scale=0.35]{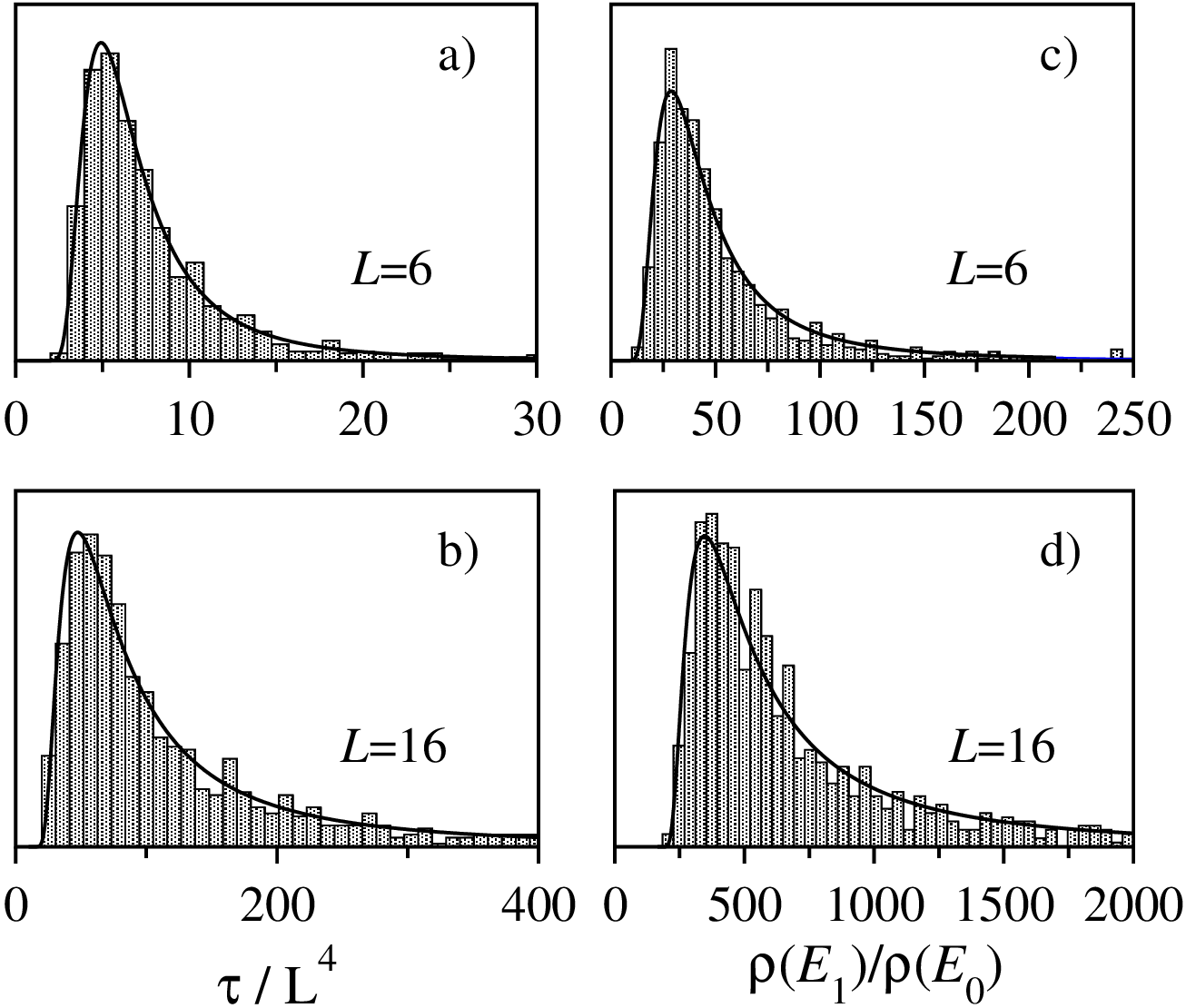}
  \caption{Normalized distributions of tunneling times (left panels) and the
    ratio of the number of first excited states to the number of ground states
    $\rho(E_1)/\rho(E_0)$ (right panels).
    For both system sizes, $L=6$ and $L=16$, 1000 randomly generated
    2D $\pm J$ spin glass realizations are sampled.
    The measured histograms of tunneling times (using local updates and
    the exact density of states) and $\rho(E_1)/\rho(E_0)$ both follow
    fat tailed Fr\'echet distributions (solid lines).}
  \label{Fig:Frechet_Distributions}
\end{figure}

\begin{figure}
  \includegraphics[scale=0.35]{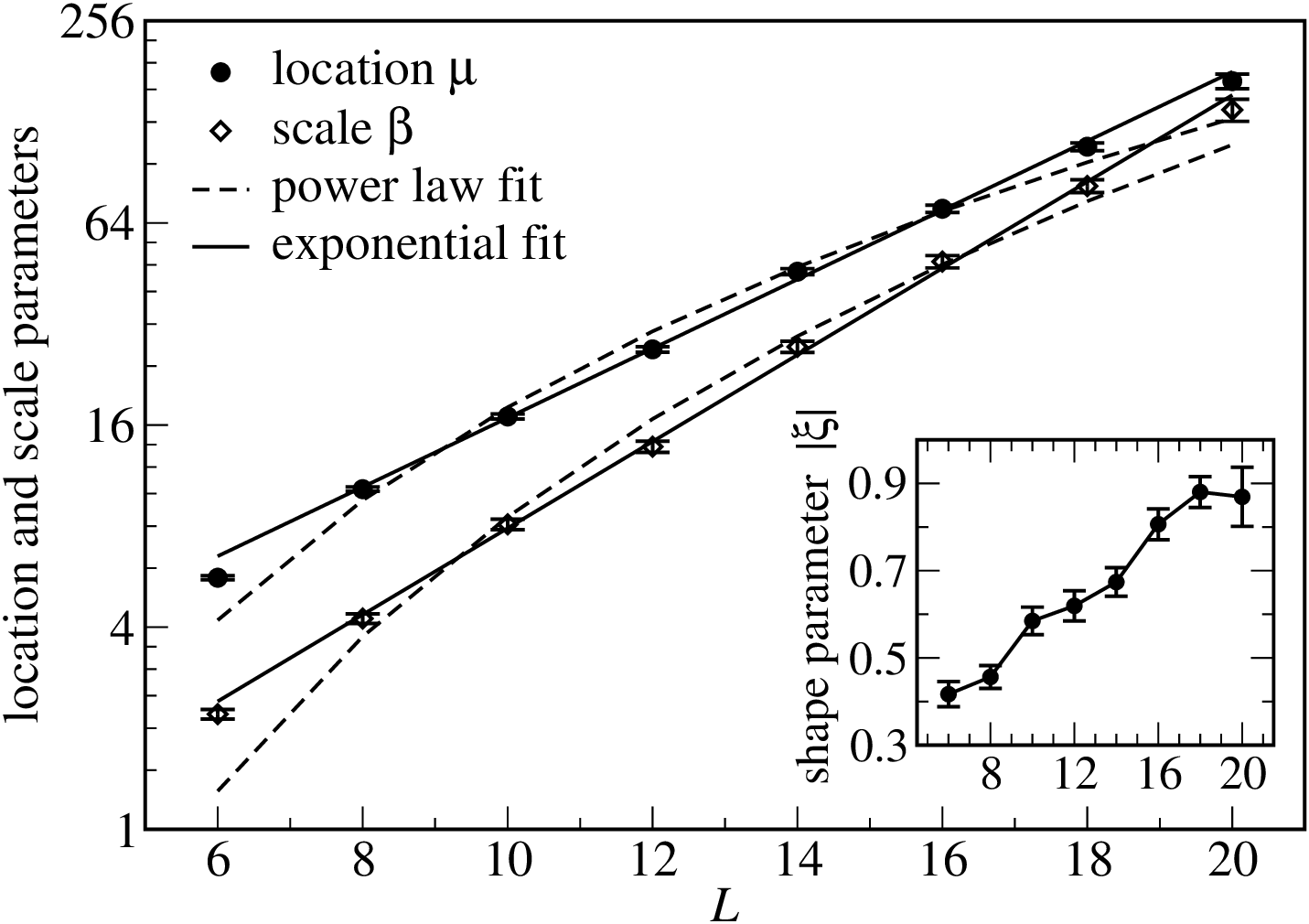}
  \caption{Scaling of the parameters of Fr\'echet distribution of the
    tunneling times (see Eq. \ref{eq:FrechetDistribution}) of the 2D $\pm J$
    Ising spin glass using perfect flat histogram sampling
    as a function of system size $L$. Solid and dashed lines
    show least square fits assuming exponential and algebraic (power law)
    scaling respectively. The exponential fits of the data with $L\ge8$ for
    $\mu$ and $\beta$  give $\chi^2=12$ and $7.6$, respectively, being significantly 
better than algebraic fits with $\chi^2=130$ and $61$,
    respectively. The inset shows the scaling of the shape parameter.}
\label{Fig:Frechet_Parameters}
\end{figure}

To determine the performance in more complex energy landscapes, we study the 2D $\pm J$ Ising spin glass, where we find exponential scaling. 
We measured tunneling times of the perfect flat histogram method for
1000 realizations for the system sizes $L=6,8,10,12,14,16,18$,
and 350 realizations for $L=20$ using the exact density of states
obtained by the algorithm of Saul and Kardar  \cite{Saul:94}.
For fixed system size the tunneling times are scattered over several orders of
magnitude for the various realizations, as shown in
Figs. \ref{Fig:Frechet_Distributions} a) and \ref{Fig:Frechet_Distributions}
b).
To analyze the underlying distributions we use extremal value theory \cite{extremal1}.
The central limit theorem for extremal values \cite{extremal2} states that the extrema
of large samples are distributed according to one of only three distributions,
depending on whether the tails of the original distribution are fat-tailed (algebraic),
exponential or thin-tailed (decaying faster than exponential). 
This theorem is successfully applied in the analysis of tails in diverse fields
such as hydrology, insurance and finance \cite{Reiss:97}. 
Surprisingly, here we find that {\it not only the extrema}, but {\it all of
  the measured tunneling times} [see Figs. \ref{Fig:Frechet_Distributions} a)
and \ref{Fig:Frechet_Distributions} b)] are distributed according to the 
Fr\'echet extremal value distribution for fat-tailed distributions:
\begin{equation}
  H_{\xi;\mu;\beta}(\tau) =
  \exp\left[ -\left( 1-\xi\frac{\tau-\mu}{\beta}\right)^{1/\xi} \right] \;,
  \label{eq:FrechetDistribution}
\end{equation}
with $\xi<0$.
The parameters of the distribution are determined by a maximum likelihood estimator.
Fig.~\ref{Fig:Frechet_Parameters} shows that
the location parameter $\mu$ specifying the maximum of the distribution
and the scale parameter $\beta$ determining the width of the distribution
scale exponentially with linear system size $L$:
\begin{eqnarray}
\mu &\propto& \exp(L/(4.21 \pm 0.04)) \;, \\
\beta &\propto& \exp(L/(3.37 \pm 0.05)) \;.
\label{eq:expfit}
\end{eqnarray}

The shape parameter $\xi$, shown in the inset of Fig.~\ref{Fig:Frechet_Parameters},
determines the power law decay of the fat tails of the distribution
\begin{equation}
  \frac{dH_{\xi;\mu;\beta}}{d\tau}
\stackrel{\tau\to\infty}{-\!\!\!\!-\!\!\!\!-\!\!\!\!-\!\!\!\!\longrightarrow}
\tau^{-(1-1/\xi)} \;.
  \label{eq:Tail}
\end{equation}
From this asymptotic behavior one can see that the $m$-th moment of a fat tailed
Fr\'echet distribution (with $\xi<0$) is well defined only if $|\xi| < 1/m$.
We find that $|\xi|>1/2$ for $L\ge10$,
which implies that the variance ($m=2$) does not exist
and the central limit theorem for mean values does not apply.
Any direct estimate of the mean tunneling time
-- as  opposed to the most likely time
given by the Fr\'echet location parameter -- then has an infinite error.
This breadth may explain differences in our conclusion from those in Refs. \onlinecite{Berg:92,Zhan:00}.
It also looks plausible, although we cannot go to large enough systems,
that $|\xi|$ increases monotonically with system size and could become larger than $1$,
in which case even the mean tunneling time ($m=1$) becomes ill-defined.
The most likely tunneling time, given by the location parameter $\mu$,
would still remain well defined and finite.


\begin{figure}
  \includegraphics[scale=0.35]{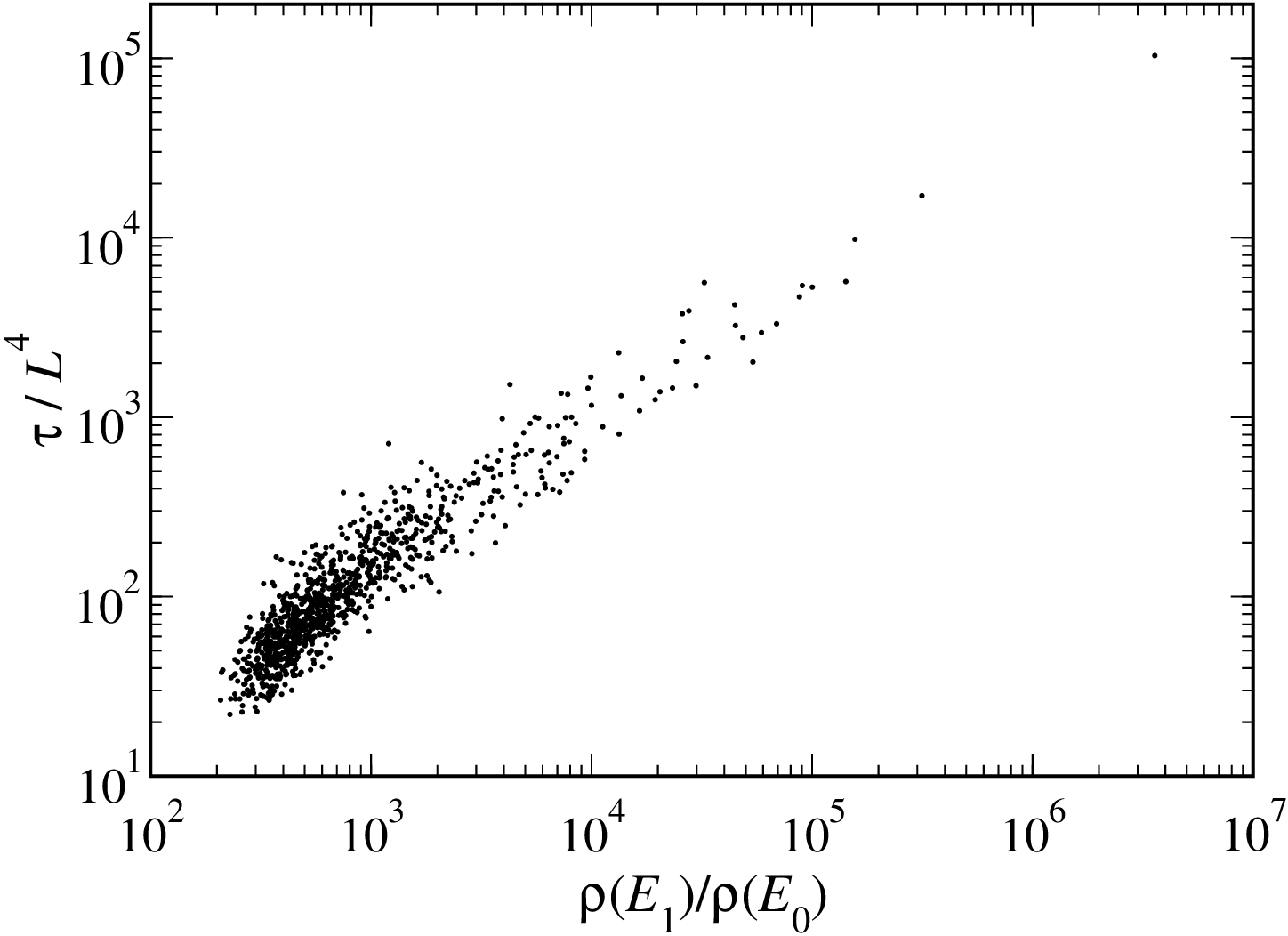}
  \caption{ Correlation between tunneling times and ratios of the number of
first excited states to the number of ground states $\rho(E_1)/\rho(E_0)$.
Shown are data from 1000 randomly 
generated 2D $\pm J$ spin glass realizations of fixed system size $L=16$. }
  \label{Fig:Correlations}
\end{figure}

Since the tunneling time is to a large extent dominated by the energy
landscape at low energies, we study, as a qualitative measure for the
complexity of the energy landscape, the distribution of  $\rho(E_1)/\rho(E_0)$
where $\rho(E_1)$ is the number of first excited states and $\rho(E_0)$ the
number of ground states.
Again we find fat tailed Fr\'echet distributions, as
shown in Figs. \ref{Fig:Frechet_Distributions} c) and
\ref{Fig:Frechet_Distributions} d), suggesting that intrinsic properties of the
2D $\pm J$ spin glass account for the observed distribution of the measured tunneling times. 
In Fig.~\ref{Fig:Correlations} we show the measured tunneling times versus the ratio 
$\rho(E_1)/\rho(E_0)$.
A strong correlation over five orders of magnitude is found, supporting the argument.
 
The question arises why the scaling behaviors of the fully frustrated model
and the spin glass are different.
Both models have an exponentially large number of ground states and an extensive
ground state entropy, but
the tunneling time scaling is algebraic for the fully frustrated model
and exponential for the spin glass.
We believe that the reason is the difference in
complexity of the energy landscapes in the two models.
While the energy landscape above the large number of ground states
of a frustrated model can be simple, the energy landscape of the spin glass is more complex
with an extremely large number of local minima:
The number of first excited states $\rho(E_1)$ that can be reached from the
$\rho(E_0)$ ground states by one single spin flip is at most $L^d\rho(E_0)$.
The number of local minima that are not connected to the ground state by
a single spin flip can thus be estimated as $\rho(E_1)-L^d\rho(E_0)$.
For the 2D ferromagnetic model the ratio $\rho(E_1)/\rho(E_0)$ is exactly
$L^2$, and $\rho(E_1)/\rho(E_0)$
is of order $L^2$ for the fully frustrated model.
In contrast, in the 2D spin glass the ratio
$\rho(E_1)/\rho(E_0)$ exceeds $L^2$ by several orders of magnitude for some
realizations, as can be seen from Figs. \ref{Fig:Frechet_Distributions} c),
\ref{Fig:Frechet_Distributions} d) and \ref{Fig:Correlations}.


\begin{figure}
  \includegraphics[scale=0.325]{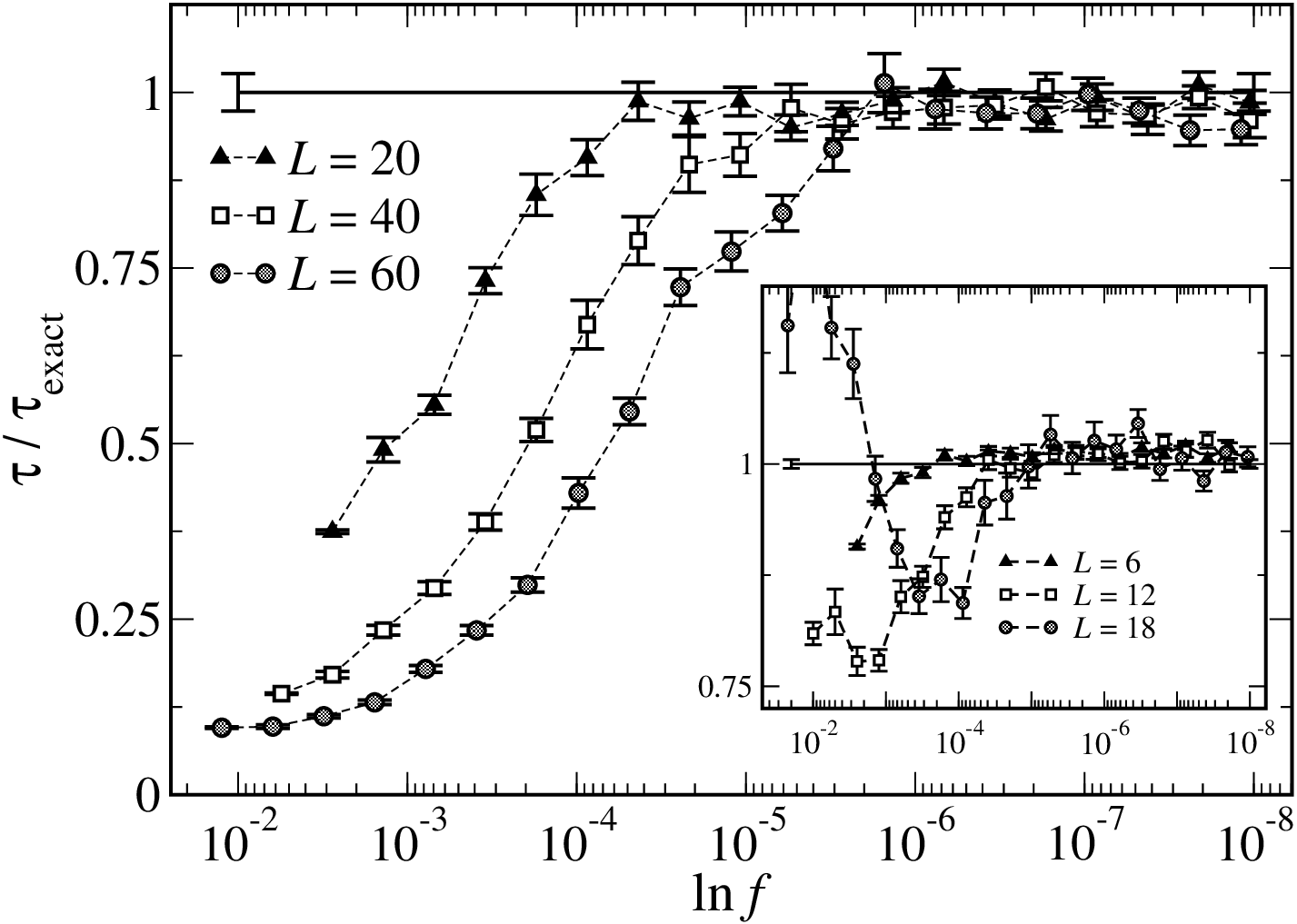}
  \caption{Convergence of the tunneling times $\tau$ during the calculation of
    the density of states using Wang Landau sampling versus Wang-Landau parameter $f$.
    Main panel: 2D Ising ferromagnet model; inset: typical samples for the 2D $\pm J$ spin glass.
    For each system size the tunneling times are averaged
    over $50$ (FM) and $500$ (SG) independent runs,
    and are given in units of the average tunneling time
    $\tau_{\rm exact}$  measured during random walks in the exact density of states
    using Wang-Landau (local update) sampling. }
  \label{Fig:Convergence_TunnelingTime}
\end{figure}

Finally we compare the tunneling times measured in the Wang-Landau algorithm
to the perfect flat histogram method.
The Wang-Landau algorithm approaches the exact density of states $\rho(E)$
by multiplying the current estimate at each visited level by a factor $f$
that is reduced towards 1 over time as the algorithm converges.
In Fig.~\ref{Fig:Convergence_TunnelingTime} we show tunneling times
for Wang-Landau sampling as a function of this correction factor $f$
for the Ising ferromagnet (main panel) and for the spin glass (inset).
Results for the fully frustrated model (not shown) are qualitatively similar.
In the initial stages of the simulation ($\ln f \gtrsim 10^{-6}$)
the tunneling times are shorter than for exact sampling,
since the random walk is biased --- it is
always driven away from the last region visited
(due to the increased $\rho(E)$ there).
Eventually the tunneling times converge to exactly the same times as for
the perfect flat histogram method, indicating convergence of the Wang-Landau algorithm.
The Wang-Landau algorithm is thus optimal in the sense that it performs identically
to a perfect flat histogram method.
Unlike in recent applications to continuum systems \cite{Yan:03}
no convergence problems are observed for these lattice models.


Our benchmarks of a perfect flat histogram method provide a lower bound for
the tunneling times of other "flat histogram" methods such as  multicanonical \cite{Multicanonical}, tempering \cite{Tempering}, broad histograms \cite{Oliveira:96} or
Wang-Landau sampling \cite{WangLandau}. From our analysis of tunneling times we find that the Wang Landau algorithm scales identically as the perfect flat histogram method and is thus optimal. We expect that the other methods will perform similarly when well-tuned.
However, whether one uses local or $N$-fold way updates, the scaling is not the $N^2$ scaling of a random walk in a system with $N$ energy levels, but slower, namely $N^2L^z$ for both the ferromagnetic ($z=0.743\pm0.007$) and the fully frustrated Ising model ($z=1.727\pm0.004$). 
The power law scaling for the frustrated model is very encouraging and demonstrates,
for the first time, that the Wang-Landau algorithm is well suited for frustrated models.
A combination with alternative sampling schemes, such as in Ref. \cite{Yamaguchi:02},
can further improve the performance.
The observation of a power law with an exponent larger than 2 is not trivially explained in the context of a random walk in energy space and the subject of further investigations.

The exponential scaling for the $\pm J$ spin glass even for the perfect method
shows a limitation for any flat histogram method. Here the distribution of
tunneling times follows a fat tailed Fr\'echet extremal value
distribution. The origin of this extremal character of the 2D $\pm J$ Ising spin glass
remains an interesting open question.
Further studies are in progress to investigate this issue as well as three-dimensional
classical spin and quantum spin glasses. 


We thank F. Alet, N. Kawashima, W. Krauth, M. Pelikan, A. Sandvik, and
B. J. Schulz for helpful discussions and acknowledge the support of the Aspen
Center for Physics where this project was started.
PD, ST, and SW acknowledge support by the Swiss National Science Foundation,
and SC and SS acknowledge support from the Petroleum Research Fund of the
American Chemical Society and by NSF-DMR.

\bibliography{paper}

\begin{thebibliography}{28}
\bibitem{footnote}
\bibnamefont{We have checked that the behavior of $\tau$ is not sensitive to details
of its definition.}
\end{thebibliography}
\end{document}